\documentclass[aps,prl,twocolumn,floatfix,amsmath,superscriptaddress,amssymb,bibnotes]{revtex4-1}

\usepackage{graphicx}
\usepackage{bm}
\usepackage{epstopdf}
\usepackage[]{SIunits}
\usepackage{units}
\addunit{\gauss}{G}

\usepackage{braket}
\usepackage{pbox}
\usepackage[colorlinks, linkcolor = blue, citecolor = blue, filecolor = black, urlcolor = blue]{hyperref}

\usepackage{xspace} 

\begin{document}

\title{Single-photon absorber based on strongly interacting Rydberg atoms}

\author{C. Tresp}
\email[]{c.tresp@physik.uni-stuttgart.de}
\affiliation{5. Phys. Inst. and Center for Integrated Quantum Science and Technology, Universit\"{a}t Stuttgart, Pfaffenwaldring 57, 70569 Stuttgart, Germany}
\author{C. Zimmer}
\affiliation{5. Phys. Inst. and Center for Integrated Quantum Science and Technology, Universit\"{a}t Stuttgart, Pfaffenwaldring 57, 70569 Stuttgart, Germany}
\author{I. Mirgorodskiy}
\affiliation{5. Phys. Inst. and Center for Integrated Quantum Science and Technology, Universit\"{a}t Stuttgart, Pfaffenwaldring 57, 70569 Stuttgart, Germany}
\author{H. Gorniaczyk}
\affiliation{5. Phys. Inst. and Center for Integrated Quantum Science and Technology, Universit\"{a}t Stuttgart, Pfaffenwaldring 57, 70569 Stuttgart, Germany}
\author{A. Paris-Mandoki}
\affiliation{5. Phys. Inst. and Center for Integrated Quantum Science and Technology, Universit\"{a}t Stuttgart, Pfaffenwaldring 57, 70569 Stuttgart, Germany}
\author{S. Hofferberth}
\email[]{s.hofferberth@physik.uni-stuttgart.de}
\affiliation{5. Phys. Inst. and Center for Integrated Quantum Science and Technology, Universit\"{a}t Stuttgart, Pfaffenwaldring 57, 70569 Stuttgart, Germany}
\date{\today}

\begin{abstract}
We report on the realization of a free-space single-photon absorber, which deterministically absorbs exactly one photon from an input pulse. Our scheme is based on the saturation of an optically thick medium by a single photon due to Rydberg blockade. By converting one absorbed input photon into a stationary Rydberg excitation, decoupled from the light fields through fast engineered dephasing, we blockade the full atomic cloud and change our optical medium from opaque to transparent. We show that this results in the subtraction of one photon from the input pulse over a wide range of input photon numbers. We investigate the change of the pulse shape and temporal photon statistics of the transmitted light pulses for different input photon numbers and compare the results to simulations. Based on the experimental results, we discuss the applicability of our single-photon absorber for number resolved photon detection schemes or quantum gate operations.

\end{abstract}

\maketitle
The elementary operation of subtracting exactly one photon from an arbitrary light pulse is of great interest for testing fundamental concepts of quantum optics \cite{Bellini2007,Lvovsky2013} as well as for the preparation of non-classical states of light for quantum information \cite{Milburn1989,Cerf2005,Polzik2006,Sasaki2010}, simulation \cite{Lukin2008,Ciuti2013,Lukin2014b}, and metrology protocols \cite{Treps2014}. Heralded single-photon subtraction has been realized by monitoring the weak reflection of a highly imbalanced beam splitter, where a single detection event corresponds to subtraction of a photon from the transmitted pulse \cite{Grangier2006,Bellini2007}. For sufficiently low reflectivity such that the subtraction of two or more photons becomes negligible, this procedure implements the photon annihilation operator $\hat{\alpha}$ \cite{Bellini2007}. This operation is inherently probabilistic, with the success rate depending on the number of incoming photons.
In contrast, deterministic single-photon subtraction, where always exactly one photon is removed independent of the input photon state, can be implemented by sending the light through a medium saturable by a single absorption event. One realization of such a single-photon absorber is a single 3-level quantum emitter strongly coupled to an optical resonator \cite{Imamoglu2008,Wilson2013}, as recently demonstrated by Rosenblum et al. using a single atom coupled to a microsphere resonator \cite{Dayan2015}.

Here we demonstrate the experimental realiziation of a deterministic free-space single-photon absorber \cite{Buechler2011}, which is based on the saturation of an optically thick free-space medium by a single photon due to Rydberg blockade \cite{Lukin2001c}. Single-photon subtraction adds a new component to the growing Rydberg quantum optics toolbox \cite{Adams2010,Vuletic2012,Vuletic2013b,Hofferberth2016d}, which already contains photonic logic building-blocks such as single-photon sources \cite{Kuzmich2012b}, switches \cite{Duerr2014}, transistors \cite{Hofferberth2014,Rempe2014b,Hofferberth2016}, and conditional $\pi$-phase shifts \cite{Duerr2016}. Our approach is scalable to multiple cascaded absorbers, essential for preparation of non-classical light states for quantum information and metrology applications \cite{Grangier2006,Polzik2006,Bellini2008}, and, in combination with the single-photon transistor, high-fidelity number-resolved photon detection \cite{Buechler2011,Weidemueller2012,Lesanovsky2011}.

Any process which deterministically removes the first photon from a light pulse will result in distortion of the pulse, reducing the purity of the output photon state \cite{Pohl2013}. While this is unproblematic for applications such as number-resolved photon detection, it imposes limits on the fidelity of photonic quantum state preparation based on photon subtraction \cite{Grangier2006,Polzik2006,Bellini2008}. We investigate the effect of the singe-photon absorption by analyzing the pulse shape and the photon-photon correlations of the output pulse. A specific feature of our system is that the photon absorption probability can be tuned via the control light parameters, enabling operation as either probabilistic or deterministic single-photon absorber. In particular, by adapting the absorption probability, the purity of the output state can be maximized for a specified input state.

\begin{figure}[t]
\includegraphics{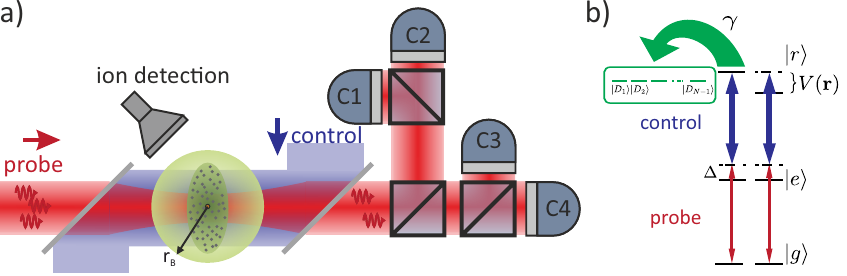}
\caption{\label{fig:experimentscheme} (a) Sketch of the free-space single-photon absorber setup. A weak probe field at $\unit[780]{\nano\metre}$ and a strong coupling field at $\unit[480]{\nano\metre}$ are overlapped with a small cloud of $^{87}\text{Rb}$ atoms coupling the $|g\rangle = |5S_{1/2},F=2,m_F=2\rangle$ ground state to the Rydberg-state $|r\rangle = |121S_{1/2},m_J=1/2\rangle$. The transmission of probe photons through the medium is monitored using four single-photon detectors (C1 to C4) in two Hanbury Brown and Twiss setups. The presence of a Rydberg excitation can be probed by field ionization and subsequent detection of $\text{Rb}^+$-ions on an ion detector (MCP). (b): Level scheme of our single-photon absorber. The two-photon excitation scheme with intermediate state detuning $\Delta = 2\pi\cdot\unit[100]\mega\hertz$ leads to strong coupling of the ground state $|g\rangle$ and the Rydberg state $|r\rangle$. If a Rydberg atom is present in the medium, strong Rydberg-Rydberg interaction $V(\textbf{r})$ prohibits subsequent Rydberg excitations and therefore absorption of probe photons. Dephasing $\gamma$ into the  many-atom dark-states results in a single Rydberg excitation decoupled from the probe light.}
\end{figure}
Our single-photon absorber scheme is based on an optically thick ensemble of 2-level emitters in free space. The complete medium, containing $N \sim 25000$ emitters, is saturated by a single photon because of strong long-range interaction between emitters in the excited state. For this purpose, we couple a weak probe light via a classical control field to realize a two-photon transition from the ground state $|g\rangle$ to a high lying Rydberg state $|r\rangle$ in an ultracold atomic ensemble. By reducing the size of the medium below the Rydberg blockade diameter \cite{Lukin2001c}, only a single Rydberg atom can be excited at a time. Because of the indistinguishability of the $N$ atoms the excited state is the single symmetric bright state $|\mathcal{W}\rangle = \frac{1}{\sqrt{N}}\sum_{i}^{}{|g_1,g_2,...,r_i,...,g_N \rangle}$, while $N-1$ many-atom dark states $|\mathcal{D}_j\rangle$ are decoupled from the light. In absence of dephasing the ensemble would undergo Rabi oscillations between the many-body states $|\mathcal{G}\rangle= {|g_1,...,g_N \rangle}$ and $|\mathcal{W}\rangle$ with collective Rabi frequency $\Omega_\text{N}$ enhanced by $\sqrt{N}$ \cite{Kuzmich2012c,Walker2014,Ott2015,Gross2015c}. In contrast, an inhomogeneous dephasing $\gamma$ acting on each atom individually will drive the system into an incoherent mixture of the bright state $|\mathcal{W}\rangle$ and all dark states $|\mathcal{D}_j\rangle$, resulting in one Rydberg excitation in the system decoupled from the coherent collective evolution \cite{Buechler2011}. This process converts exactly one photon of the probe field into a stationary Rydberg atom, changing the medium from opaque to transparent for subsequent photons due to the Rydberg interaction induced level shifts.

Experimentally, we implement the absorber by focusing a weak $\unit[780]{\nano\metre}$ probe beam (waist $w_{0,\text{probe}} = \unit[6.5]\micro\metre$) into a cold ($T=\unit[8]\micro\kelvin$) atomic cloud  ($\sigma_z=\unit[6]\micro\metre$, $\sigma_r=\unit[10]\micro\metre$) containing $25000$ optically trapped $^{87}\text{Rb}$ atoms (Fig.\,\ref{fig:experimentscheme} a). In this geometry we measure an optical depth of $OD_b = 12.5$ when scanning the frequency of the probe field over the transition from the initial state $|g\rangle = |5S_{1/2},F=2,m_F=2\rangle$ to the intermediate state $|e\rangle = |5P_{3/2},F=3,m_F=3\rangle$. We couple the probe photons to the Rydberg state $|r\rangle = |121 S_{1/2},m_J=1/2\rangle$ via a strong control field at $\unit[480]\nano\metre$ (beam waist $w_{0,\text{control}} = \unit[14]\micro\metre$). In the following experiments, both lasers are detuned from the intermediate state by $\Delta = 2\pi\cdot\unit[100]\mega\hertz$, while the two-photon detuning $\delta=0$ so that the combined fields are in resonance with the $|g\rangle \rightarrow |r\rangle$ transition (Fig.\,\ref{fig:experimentscheme} b). The calculated, spatially averaged probe Rabi frequency is $\Omega_p = 2\pi \cdot 33 \sqrt{\frac{\text{photons}}{\unit[]\micro\second}} \unit[]\kilo\hertz$ while the control Rabi frequency is measured to be $\Omega_c =2\pi\cdot \unit[10]\mega\hertz$. The resulting Rydberg blockade radius $r_B\approx\unit[17]\micro\metre$ significantly exceeds the size of the atomic cloud.

To study the saturation of our medium, we send Tukey-shaped input probe pulses with mean photon number $\overline{N}_{in}$ and duration $\tau\approx\unit[2]\micro\second$ through the medium and measure the mean number of transmitted photons $\overline{N}_{out}$ using four avalanche single-photon detectors (SPCM) in two Hanbury Brown and Twiss setups (HBT), as shown in Fig.\,\ref{fig:experimentscheme} a. The control light is kept constant over the duration of the full probe pulse. After each pulse, we probe the presence of a Rydberg atom by applying a field ionization pulse to convert any Rydberg atoms into $\text{Rb}^+$ ions which are subsequently detected on a microchannel plate (MCP).

First, we record the absorption spectrum of the probe for small mean photon number $\overline{N}_{in} = 0.2$ to determine the the residual probability $p_\text{scatt}=0.01$ to scatter a probe photon from the intermediate state and the probability $p_\text{Ryd} = 0.35$ to convert a single probe photon into a Rydberg excitation by measuring with the control light off and on, respectively. Since for $\overline{N}_{in} \ll 1$ there is no saturation, we can extract the effective dephasing rate $\gamma = 2\pi\cdot\unit[500]\kilo\hertz$ by fitting the observed Rydberg absorption line with the well-known solution of the optical Bloch equations for the three-level atom \cite{Fleischhauer2005}, which include the Rydberg state lifetime $\tau_\text{Ryd} = \unit[530]\micro\second$ and the spontaneous Raman decay rate $\gamma_\text{Raman} = \left(\frac{\Omega_c}{2\Delta}\right)^{2} \Gamma_\text{5P}$, where $\Gamma_\text{5P} =2\pi \cdot \unit[6.05]\mega\hertz$ is the intermediate state spontaneous decay rate.
We consider three main contributions to the observed dephasing rate $\gamma$. First, thermal motion of the atoms results in intrisic dephasing proportional to the atomic velocity \cite{Pan2009,Kuzmich2013b}. Secondly, we perform all experiments in-trap to induce a spatially varying shift of the two-photon resonance due to the trap-induced ac-Stark effect, which is equivalent to an inhomogeneous dephasing \cite{Buechler2011}. Finally, elastic collisions of the Rydberg electron with ground state atoms result in a dephasing proportional to the atomic density \cite{Duerr2014,Pfau2014c}. From the measured temperature and density of our atomic cloud and the optical trap depth we expect all three effects to contribue dephasing rates of order $\sim 2\pi \unit[100]\kilo\hertz$, which agrees well with the measured effective dephasing $\gamma = 2\pi\cdot\unit[500]\kilo\hertz$.

The observed dephasing and decay rates determine the optimal probe pulse duration and collective Rabi frequency $\Omega_\text{N}$. The pulse should be longer than $1/\gamma$, while $\Omega_\text{N} \approx \gamma$, to ensure that the collective excitation reliably dephases into a stationary Rydberg atom during the probe pulse duration \cite{Buechler2011}. At the same time, the pulse must be short compared to the Rydberg atom lifetime, so that the initial excitation blocks the medium for the full pulse duration. In our case the Rydberg lifetime is set by the spontaneous Raman decay rate $\gamma_\text{Raman}$ due to admixture of the intermediate state by the detuned control field. Spontaneous or black-body radiation induced transitions as well as inelastic collisions with ground state atoms result in much smaller decay rates for $|r\rangle = |121 S_{1/2}\rangle$ and our atomic density \cite{Ott2015b,Pfau2016}.

\begin{figure}
\includegraphics{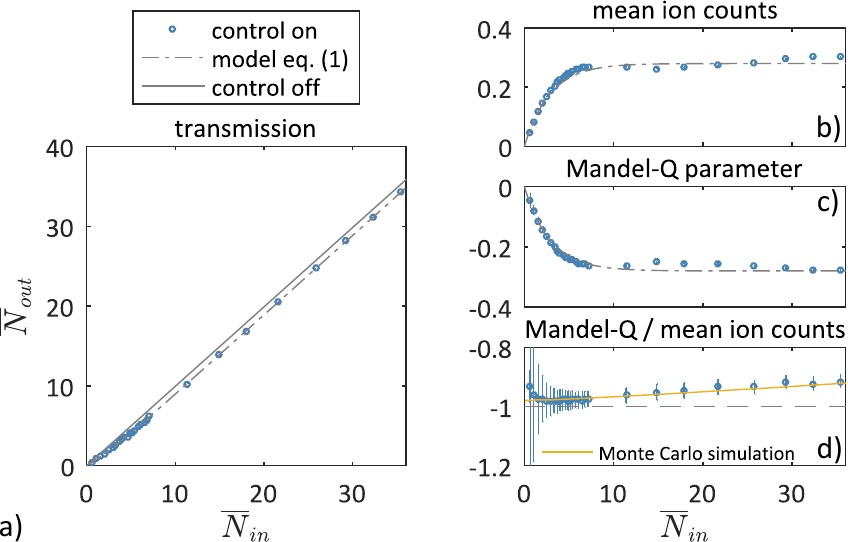}
\caption{\label{fig:nonlinearity} (a): The number of transmitted photons $\overline{N}_{out}$ is plotted against the number of input $\overline{N}_{in}$ (circles). Each datapoint is the average of 250000 experiment repetitions. The solid line shows the measured linear transmission when the control field is off limited by the finite transmission $t=0.99$ of the probe field at a detuning of $\Delta = 2\pi\cdot\unit[100]\mega\hertz$ . The dash-dotted line is the result of our simulation assuming Poissonian statistics of the input light and the measured single-photon absorption probability eq. \ref{eq:nonlinearity}. The number of transmitted photons is reduced exactly by one over a large range of input photon numbers. (b-d): Statistics of Rydberg atom detection via field-ionization. Saturation of the mean ion number and negative Mandel-Q parameter reveal full blockade of the atomic medium by a single Rydberg excitation. Error bars show the standard error of the mean and are smaller than the markers where not visible.}
\end{figure}

Fig.\,\ref{fig:nonlinearity}a shows the mean number of transmitted probe photons $\overline{N}_{out}$ for different input photon numbers $\overline{N}_{in}$ per probe photon pulse. When the control field is off the probe transmission $t=1-p_\text{scatt}=0.99$ is independent of the input photon number (solid line). In contrast, with the control light on (circles), we observe a strongly nonlinear behaviour. For small probe input $\overline{N}_{in}<10$, the number of absorbed photons is determined by the Poissonian statistics of the coherent probe field and our finite Rydberg absorption probability $p_\text{Ryd} $. For $\overline{N}_{in}>10$, we observe that the mean number of transmitted photons is  reduced by one, independent of the input photon number ($\Delta \overline{N} = 0.98(13)$) averaged over all data with $\overline{N}_{in}>10$). The dashed line shows the behaviour of a single-photon absorber with $p_\text{Ryd} = 0.35$ and $t=0.99$, but otherwise ideal performance, given by the expression
\begin{equation}
\overline{N}_{out} = t\cdot\overline{N}_{in}+\text{exp}\left(-t\cdot\overline{N}_{in}\cdot p_\text{Ryd}\right)-1,
\label{eq:nonlinearity}
\end{equation}
which agrees very well with our data, suggesting that our medium is indeed saturated by a single absorption event. To further substantiate this claim, we show the corresponding ion statistics recorded with the MCP in Fig.\,\ref{fig:nonlinearity}b-d. For increasing probe input, the mean ion count per probe pulse saturates at $0.29$ (Fig.\,\ref{fig:nonlinearity}b), limited by the detection efficiency $\eta$ of our MCP. For clear evidence of the blockade in our medium, we calculate the Mandel-Q parameter $Q  = \frac{\text{Var}(n)}{\langle n\rangle}-1$, where $n$ is the measured distribution of single shot ion counts (Fig.\,\ref{fig:nonlinearity}c). Again, the experimental data saturates at $Q = -\eta$, which is the expected limit if there is exactly one ion created in each run, caused by the falsified number of zero events due to finite detection efficiency. The observed mean ion count and the Mandel Q-values both agree well with the expected results based on Eq.\,\ref{eq:nonlinearity}, assuming that each absorbed photon is subsequently converted into an ion and detected with $\eta=0.29$.
Finally, we calculate the ratio of Mandel-Q parameter to mean number of detected ions (Fig.\,\ref{fig:nonlinearity}d), which should be constant at -1 for perfect Rydberg blockade. Instead, we observe a slight increase from -0.98 at $\overline{N}_{in} = 3$ to -0.91 at $\overline{N}_{in} = 35$. This effect cannot be explained within our simple analytic model (Eq. \ref{eq:nonlinearity}) allowing only absorbtion of a single photon. We thus numerically simulate the absorption process including the possibility that a second photon is absorbed even if the medium is already blockaded. To do so, we consider pulses divided into bins of $\text{t}_\text{bin} = \unit[50]\nano\second$, with each of the bins having a Poissonian probability to contain a certain number of photons given by the envelope function with the shape of the pulse used in the experiment and normalized to the mean input photon number. The absorption of the pulse is then simulated by going through each time bin and determining for each photon if it is absorbed by comparing a random number against the absorption probability $p_\text{Ryd}$. If one photon has already been absorbed, we allow the absorption of a second photon by instead comparing a random number to the (much smaller) probability $p_\text{Ryd2}$. From these simulations we obtain the full statistics of both the transmitted photons and of Rydberg atoms excited in the atomic medium. In particular, we find that this model reproduces the observed ratio of Mandel-Q value and mean ion number if we set $p_\text{Ryd2} = 0.001$ (orange dashed line in Fig.\,\ref{fig:nonlinearity}d). This suggests that the Rydberg blockade prevents excitation of a second Rydberg atom with very high, but not unit, fidelity. We suspect that this is due to the more complex nature of the Rydberg-Rydberg interaction than the usually assumed single van-der-Waals potential \cite{Shaffer2006,Hofferberth2015}, resulting in resonances for two-atom excitation within the conventional blockade volume \cite{Rost2007,Weidemueller2010b}. We note that the numerics including $p_\text{Ryd2} = 0.001$ result in negligible difference compared to Eq.\,\ref{eq:nonlinearity} for the quantities shown in Fig.\,\ref{fig:nonlinearity}a-c) over the shown input photon range, this effect only becomes visible by analyzing the full ion statistics.

\begin{figure}
\includegraphics{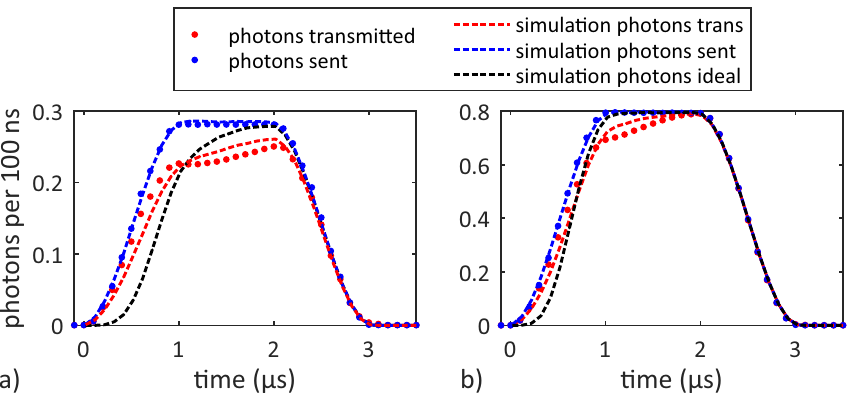}
\caption{\label{fig:pulses} Measured (points) and simulated (dashed lines) pulse shapes for $5.65$ (a) and $15.76$  (b) input photons. The change of the medium from opaque to transparent after the absorption of the first photon leads to a change in the transmitted pulse shape (red points) compared to the pulse sent into the medium (blue points). This effect is captured by our numerical simulation including single- and two-photon absorption probabilities $p_\text{Ryd} = 0.35$ and $p_\text{Ryd2} = 0.001$ (red line).  The pulse distortion is strongest for an ideal deterministic single-photon absorber ($p_\text{Ryd} = 1$ and $p_\text{Ryd2} = 0$, black line).}
\end{figure}

Having demonstrated the functionality of the single-photon absorber scheme, we next investigate the pulse shape and temporal correlations of the transmitted light. Fig.\,\ref{fig:pulses} shows the shape of input (blue dots) and output pulses (red dots) for $\overline{N}_{in}=5.65$ (a) and $\overline{N}_{in}=15.76$ (b), respectively. We observe a visible distortion of the pulse shape as photons are predominantly absorbed in the beginning of the pulse. In particular, for the pulses containing $15.76$ photons, the transmission increases to unity in the last third of the pulse, since at these times the probability that one photon has already been absorbed converges to unity. For $\overline{N}_{in} = 5.65$ photons this effect is less dramatic as the probability of all photons being transmitted is still finite ($\sim 14\%$) at this mean input. Our numerical simulation (red lines) reproduce the observed pulse shapes quite well. To show that the pulse distortion becomes more severe for higher absorption probabilities, we also show the simulated pulse shapes for the perfect single-photon absorber with $p_\text{Ryd} = 1$ and $p_\text{Ryd2} = 0 $ (black dashed lines). In this case, it is always the first photon in the pulse which is absorbed, resulting in the stronger pulse distortion at the beginning of the pulse. This observation has important consequences for different applications of the single-photon absorption scheme. For efficient number-resolved photon detection by an array of single-photon absorbers, high absorption and strong dephasing are essential \cite{Buechler2011}. In this case, one has to keep in mind that each photon subtraction results in Fourier-broadening of the pulse, which can reduce the efficiency of subsequent absorbers. In turn, for high-fidelity quantum state preparation, the pulse distortion should be minimal. In principle, the absorption in our system is tunable, enabling adapting the single-photon absorption probability such that the total absorption for a given photon number reaches unity, while the information gained about which photon is absorbed is minimal.

\begin{figure}
\includegraphics{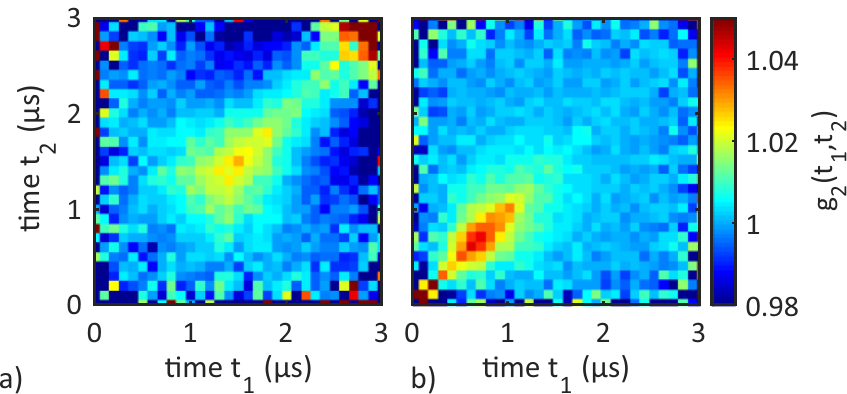}
\caption{\label{fig:correlations} (a) Measured intensity correlation function $g_2(t_1,t_2)$ for mean probe photon input $\overline{N}_{in}=15.76$. The absorption of a single photon results in photon bunching at the front of the pulse, while the photon statistics during the latter part, where the medium is transparent, are unchanged. (b) Simulated correlation functions from our numerical model qualitatively reproduce the features observed in the experimental data, but do not capture the full dynamics of the excitation and dephasing process.}
\end{figure}

Finally, to investigate the photon statistics of the transmitted light, we calculate the time dependent intensity correlation function
\begin{equation}
 g_2(t_1,t_2) = \frac{\langle n_1(t_1)\cdot n_2(t_2)\rangle}{\langle n_1(t_1)\rangle\langle n_2(t_2)\rangle},
\label{eq:correlation}
\end{equation}
where $n_1(t),n_2(t)$ are the number of detection on two different detectors at time $t$. In practice, we calculate the correlation between all distinct pairs of our four counters and average over these results. In Fig.\,\ref{fig:correlations}a we show the measured intensity correlations for mean input $\overline{N}_{in}=15.76$. We observe photon bunching, i.e. $g_2(t_1,t_2) >1$, in the time range where the single-photon absorption happens. This somewhat counter-intuitive result for an absorption process is caused by the fact that the single-photon absorber reduces the mean of the transmitted light by one, but keeps the width of the photon distribution constant, resulting in super-Poissonian photon statistics. This effect vanishes for the later part of the pulse, where we observe $g_2(t_1,t_2) = 1$, because the saturated medium no longer absorbs photons, resulting in no more modification of the photon statistics of the coherent input pulse. In Fig. \ref{fig:correlations}b we show the simulated correlation function obtained from our numerics including the finite single- and two-photon absorption probabilities. The simulation qualitatively reproduces the bunching feature and the change of the correlations over the pulse duration. The most visible difference is the time where the bunching feature appears, which stems from the fact that our numerics neglect any timescale of the initial excitation and dephasing dynamics. A more sophisticated approach, that still yields probe photon statistics, will require calculating the dynamics of the atoms in the presence of a propagating, quantized probe field.

In conclusion, we have implemented a scheme for deterministic subtraction of one photon from a probe pulse based on the fast dephasing of a single collective Rydberg excitation \cite{Buechler2011}. Because of its free-space implementation, our system is easily scalable with existing ultracold atom trapping techniques and can be combined with other Rydberg-based tools for single-photon manipulation \cite{Kuzmich2012b,Duerr2014,Hofferberth2014,Rempe2014b,Hofferberth2016,Duerr2016}. The main shortcoming of our current implementation is the limited absorption probability $p_\text{Ryd}=0.35$. This number can straightforwardly be increased, without changing any other performance aspects of our scheme, by trapping more atoms within the single blockade volume. Reaching $p_\text{Ryd}>0.8$ is possible for our current blockade without running into density limitations due to collisions between Rydberg and groundstate atoms \cite{Ott2015b,Pfau2016}. This would open the door for realization of high-fidelity number-resolved photo-detection \cite{Buechler2011}. In contrast, the ability to tune $p_\text{Ryd}$ should be beneficial for high-fidelity preparation of non-classical light states for quantum information \cite{Grangier2006,Polzik2006,Bellini2007,Bellini2008,Lvovsky2013} and metrology \cite{Sasaki2010,Treps2014}.


\begin{acknowledgments}
We thank Jan Kumlin, Przemyslaw Bienias, Hans Peter B\"{u}chler, Robert L\"{o}w, Tilman Pfau, and Yanli Zhou for fruitful discussions and comments on the manuscript. We acknowledge funding by the German Research Foundation through Emmy-Noether-grant HO 4787/1-1 and within SFB/TRR21, and by the Ministry of Science, Research and the Arts of Baden-W\"{u}rttemberg through RiSC grant 33-7533.-30-10/37/1. H.G. acknowledges support from the Carl-Zeiss Foundation.
\end{acknowledgments}

\bibliographystyle{apsrev4-1}

%

\end{document}